# SASE: Complex Event Processing over Streams

Daniel Gyllstrom[1]  Eugene Wu[2]  Hee-Jin Chae[1]
Yanlei Diao[1]  Patrick Stahlberg[1]  Gordon Anderson[1]

[1]Department of Computer Science
University of Massachusetts, Amherst
{dpg, chae, yanlei, patrick, gordon}@cs.umass.edu

[2]Computer Science Division
University of California, Berkeley
sirrice@berkeley.edu

## ABSTRACT
RFID technology is gaining adoption on an increasing scale for tracking and monitoring purposes. Wide deployments of RFID devices will soon generate an unprecedented volume of data. Emerging applications require the RFID data to be filtered and correlated for complex pattern detection and transformed to events that provide meaningful, actionable information to end applications. In this work, we design and develop SASE, a complex event processing system that performs such data-information transformations over real-time streams. We design a complex event language for specifying application logic for such transformation, devise new query processing techniques to efficiently implement the language, and develop a comprehensive system that collects, cleans, and processes RFID data for delivery of relevant, timely information as well as storing necessary data for future querying. We demonstrate an initial prototype of SASE through a real-world retail management scenario.

## 1. INTRODUCTION
Recent advances in RFID technology have facilitated its adoption in a growing number of applications including supply chain management [4], surveillance [4], and healthcare [4], to name a few. The driving force behind RFID adoption is unprecedented visibility into systems that have to this point been unobservable. With this visibility, it will be possible to monitor, correct, control, and improve processes of vital economic, societal, and environmental importance. For example, real-time visibility into supply chain inventory can help detect and prevent out-of-stocks and shrinkage before they occur. Similarly, real-time monitoring of patients taking medications can help enforce medical compliance and alert care providers when anomalies occur.

The successful development of an RFID data management system for providing such real-time visibility must address two unique challenges presented by RFID technology:

- **Logic complexity**: Data streams emanating from RFID sensing devices carry primitive information about the object a tag is attached to: its location and the time of sensing. RFID-based monitoring applications, however, require meaningful, actionable information (*e.g.*, out-of-stocks, shoplifting) that is defined by unique complex logic involving filtering, pattern matching, aggregation, recursive pattern matching, *etc*. To resolve the mismatch between data and information, it is critical to have a processing component that resides on RFID streams and performs data-information transformation.

- **Performance requirements**: Large deployments of RFID devices have the potential to create unprecedented volumes of data. Yet despite the volume of data and logic complexity, RFID data processing needs to be fast. Filtering, pattern matching, and aggregation must all be performed with low-latency.

We design and develop a complex event processing system, SASE, that transforms real-time RFID data into meaningful, actionable information. We first provide an expressive, user-friendly event language that allows applications to encode their complex logic for such data-information transformation. This language significantly extends existing languages such as complex event languages [2][9] developed for active databases and stream languages [1][3][6] with support for sequence patterns that involve temporal order of events, negation, value-based predicates, sliding windows, *etc*. It is worth noting that our event language is a general purpose language that can be applied to many application domains. We use an RFID-based scenario because it represents an important emerging class of applications that demand complex event processing and is able to demonstrate the various features of our event language.

We then provide a query plan-based approach to efficiently implementing the proposed event language. This approach is based on a new abstraction of complex event processing; that is a dataflow paradigm with native sequence operators at the bottom, pipelining query-defined sequences to subsequent relational style operators. This new abstraction also allows us to explore alternative query plans to optimize for various issues in high-volume complex event processing.

We further develop a comprehensive system that collects and cleans data from RFID devices, creates events, and runs the event stream through the complex event processor to deliver timely results to the user and archive events in a database. Our system allows the user to query the resulting event database by either sending ad-hoc queries or writing continuous queries that combine stream processing and database access.

A prototype of the SASE system will be demonstrated through a simulated retail-store RFID deployment. Specifically, we demonstrate (1) the expressiveness of the language by showing its uses in specified monitoring queries as well as data transformation rules for archiving, (2) a complete data flow from RFID devices through various SASE components to final query output, and (3) track-and-trace queries over an event database.

The remainder of this paper is organized as follows. Section 2 describes our complex event language and its implementation. Section 3 presents the SASE system architecture. Section 4 discusses our demonstration scenario.

## 2. COMPLEX EVENT PROCESSING

In this section, we survey the SASE event language through examples, and discuss its implementation.

### 2.1.1 Complex Event Language

The SASE language has a high-level structure similar to SQL for ease of use by database programmers, but the design of the language is centered on event pattern matching, as illustrated in this section. The overall structure of the SASE language is:

| [FROM | <stream name>] |
|---|---|
| EVENT | <event pattern> |
| [WHERE | <qualification>] |
| [WITHIN | <window>] |
| [RETURN | <return event pattern>] |

The semantics of the language are briefly described as follows: The FROM clause provides the name of an input stream. If it is omitted, the query refers to a default system input. The EVENT, WHERE and WITHIN clauses form the event matching block. The EVENT clause specifies an event pattern to be matched against the input stream. The WHERE clause, if present, imposes value-based constraints on the events addressed by the pattern. The WITHIN clause further specifies a sliding window over the event pattern. The event matching block transforms a stream of input events to a stream of new *composite event*s.

Finally, the RETURN clause transforms the stream of composite event for final output. It can select a subset of attributes and compute aggregate values like the SELECT clause of SQL. It can also name the output stream and the type of events in the output. It can further invoke database operations for retrieval and update.

We further explain these language constructs using examples drawn from an RFID-based retail store scenario (which is used in our demonstration and is further described in Section 4). In this scenario, an RFID tag is attached to every product in a retail store. RFID readers are installed above the shelves, checkout counters, and exits. These readers generate a reading if a product is in its read range. In our examples, we assume that readings at the shelves, checkout counters, and exits are represented as events of three distinct types.

Our first example query (Q1) detects shoplifting activity; it reports items that were picked at a shelf and then taken out of the store without being checked out. The EVENT clause contains a SEQ construct that specifies a *sequence* in a particular order; the sequence consists of the occurrence of a *SHELF_READING* event followed by the non-occurrence of a *COUNTER_READING* event followed by the occurrence of an *EXIT_READING* event. Non-occurrences of events, referred to as **negation**, are expressed using '!'. For the use of subsequent clauses, the SEQ construct also includes a variable in each sequence component to refer to the corresponding event. The WHERE clause of Q1 uses these variables to form predicates that compare attributes of different events, referred to as **parameterized predicates**. The parameterized predicates in Q1 compare the *TagId* attributes of all three events in the SEQ construct for equality. Q1 contains a WITHIN clause to specify a **sliding window** over the past 12 hours. Finally, the RETURN clause retrieves the *TagId* and *ProductName* of the item, the *AreaId* of the exit, and initiates a database lookup to retrieve a textual description of the exit (*e.g.*, the leftmost door on the south side of the store). Note that our language provides a set of built-in functions (all starting with '_') for common database operations and can be extended to accommodate other user functions.

Q1: EVENT    SEQ(SHELF_READING x, ! (COUNTER_READING y), EXIT_READING z)
     WHERE    x.TagId = y.TagId $\wedge$ x.TagId = z.TagId
     WITHIN    12 hours
     RETURN    x.TagId, x.ProductName, z.AreaId,
                   _retrieveLocation(z.AreaId)

The second example (Q2) illustrates the use of the SASE language to express data transformation rules for archiving. Here, we use an event sequence query to detect a change in location of an item and trigger a database update to reflect the change. The EVENT, WHERE, and WITHIN clauses are used to detect the location change. The RETURN clause calls a system function *_updateLocation* to perform a location update in the database. Internally, the event database stores the location of an item using *TimeIn* and *TimeOut* attributes, representing the duration of its stay. The *_updateLocation* function first sets the *TimeOut* attribute of the current location using the *y.Timestamp* value, and then creates a tuple for the new location with the *TimeIn* attribute also set to the value of *y.Timestamp*.

Q2: EVENT    SEQ(SHELF_READING x, SHELF_READING y)
     WHERE    x.id = y.id $\wedge$ x.area_id != y.area_id
     WITHIN    1 hour
     RETURN    _updateLocation(y.TagId, y.AreaId, y.Timestamp)

Several other queries supported by our language will be shown through our demonstration, as described in Section 4.

### 2.1.2 Implementation

SASE is implemented using a query plan-based approach, that is, a dataflow paradigm with pipelined operators as in relational query processing. As such, it provides flexibility in query execution and extensibility as the event language evolves. This approach employs a new abstraction for event query processing. Specifically, we devise native sequence operators based on a *Non-deterministic Finite Automata* based model which can read query-specific event sequences efficiently from continuously arriving events. These operators are then used to form the foundation of each plan, pipelining the event sequences to subsequent operators such as selection, window, negation, *etc*. This arrangement allows the subsequent operators to be implemented by leveraging relational techniques.

The new abstraction of event query processing also allows us to optimize for two salient issues in complex event processing: *large sliding windows* and *intermediate result sets*. Large sliding windows spanning hours or days are commonly used in monitoring applications. Sequence generation from events widely dispersed in such windows can be an expensive opera-tion. To address this issue, we develop optimizations that employ novel sequence indexes to expedite the sequence operators. Large intermediate result sets also strongly affect query processing. To reduce intermediate results, we strategically push some of the predicates and windows down to the sequence operators; the optimizations are based on indexing relevant events both in temporal order and across value-based partitions. The interested reader is referred to [9] for details of these techniques.

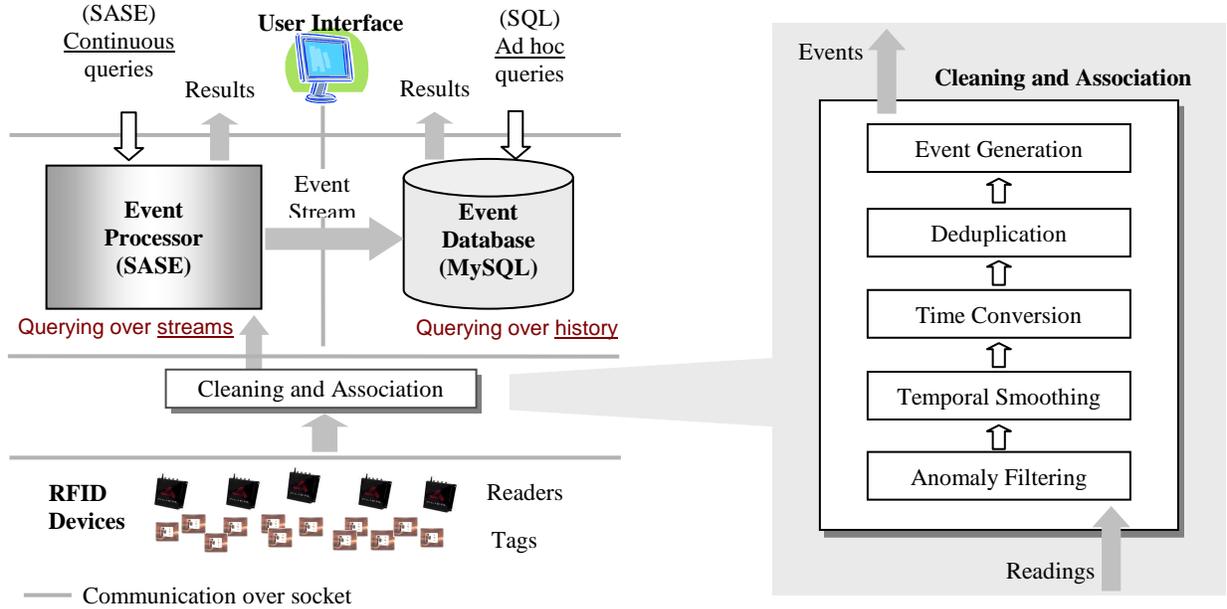

Figure 1: SASE System Architecture

## 3. ARCHITECTURE

The architecture of the SASE system is shown in Figure 1. It consists of three layers. The bottom layer contains physical RFID devices (*e.g.* tags, readers). The RFID data returned from RFID readers is passed to the next layer for data cleaning and event generation. The event stream is then fed to the third layer where most of the data processing takes place. A core component of the layer is a complex event processor that processes the event stream to deliver timely results to the user and archive events into a database. Our system allows the user to query the resulting event database by either sending ad-hoc queries or writing continuous queries that combine stream processing and database access. These components are described in more detail below.

**Physical Device Layer**. The physical device layer consists of RFID readers, antennas, and tags. RFID readers scan their reading range in regular intervals and return a reading for each detected tag. Each raw RFID reading consists of the *TagId* and *ReaderId*. For our demonstration, we use a Mercury 4 Agile RFID Reader from *Thingmagic* and multiple antennas to simulate multiple readers. Individual objects are tagged with EPC Class1 Generation 1 tags from Alien Technology.

**Cleaning and Association Layer**. The Cleaning and Association Layer serves two important functions. First, it copes with idiosyncrasies of readers and performs data cleaning, such as filtering and smoothing. This is important as RFID readings are known to be inaccurate and lossy. Our data cleaning component is based on some of the techniques described in [8]. Second, it uses attributes such as product name, expiration date, and saleable state to create events. This helps facilitate processing and decision making in subsequent components. Internally, the Cleaning and Association layer consists of five components:

1) Anomaly Filtering Layer: removes spurious readings and readings that contain truncated ids.
2) Temporal Smoothing Layer: the system decides whether an object was present at time *t* based not only on the reading at time *t*, but also on the readings of this object in a window size of *w* before *t*. Using this heuristic, a new reading may be created.
3) Time Conversion Layer: a timestamp is appended to each reading based on a logical time unit that is set as a system configuration parameter.
4) Deduplication Layer: removes duplicates, which can be caused either by a redundant setup, where two readers monitor the same logical area, or when an item resides in overlapping read ranges of two separate readers.
5) Event Generation Layer: generates events according to a pre-defined schema. An important step in event generation is to obtain attributes defined in the schema. In an actual real-world system, attributes (*e.g.*, product name, expiration date) can be retrieved from a tag's user-memory bank or from an Object Name Service (ONS) [5]. In our system, we simulate an ONS with a local database storing product metadata associated with each item.

**Complex Event Processor**. The complex event processor supports continuous long-running queries written in the SASE language over event streams. Specifically, it performs three tasks.

- For each monitoring task such as detection of shoplifting, the user writes a query and registers it as a *continuous query* with the complex event processor. The event processor immediately starts executing the query over the RFID stream and returns a result (*e.g.*, a notification) to the user every time the query is satisfied. Such processing continues until the query is deleted by the user.
- Transformation rules for data archiving are also registered as continuous queries with the event processor. These queries can be used to remove duplicate data and transform data to the format required for archival. The resulting events are streamed to the event database for storage.
- The event processor can further handle complex continuous queries that integrate stream processing and database lookup; upon detection of an event of interest, these queries require database access to retrieve additional information, as shown in Q1 in Section 2.1.1. The event processor supports these queries by first detecting the event, then sending a subquery to the database, combining information retrieved from the database with that obtained from the stream, and finally returning a complete result to the user.

**Event Database**. SASE contains a persistence storage component to support querying over historical data and to allow query results from the stream processor to be joined with stored data. We use MySQL 5.0.22 as our DBMS. As mentioned in the previous section, RFID stream data is transformed using rules declared with the complex event processor before archiving. Our system supports two important rules: Location Update and Containment Update. For location update, a tag's location information is updated when we observe this tag in a different location with a different timestamp. For containment updates, readings from unloading and loading zones are aggregated into a containment relationship.

**User Interface (UI)**. SASE has a UI that allows the user to issue both continuous queries over the RFID stream and ad hoc queries on the event database. It provides a visual presentation for the query results as well as the internal data flow through various SASE system components. The UI provides separate windows for monitoring the events output by the Cleaning and Association Layer, presenting stream processor query results, and displaying results from event database queries.

## 4. DEMONSTRATION SCENARIO

Our demonstration is based on a simulation of a typical retail management scenario. The retail store setup consists of four readers (antennas), with one reader in each of the following locations: the store exit, two shelves, and check-out counter. Each reader occupies only one logical area. This setup is depicted in Figure 2.

Using this setup, we first have a live demonstration where actions (e.g. shoplifting or misplaced inventory) are simulated live in our retail-store setup. Continuous queries monitor and detect these actions. Then we execute track-and-trace queries over an event database populated with data collected in advance.

Through our demonstration, we show
1) The expressiveness of the SASE language through its uses for monitoring queries and data transformation rules for archiving.
2) The transformation of raw RFID data streams to a semantic level appropriate for higher level monitoring applications.
3) The actions performed by each of the SASE components by showing a complete data flow from RFID devices through each of the SASE components to final query output.

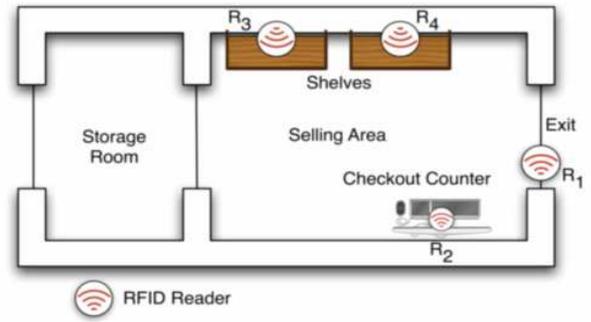

Figure 2: Demonstration Setup

4) Track-and-trace queries over the event database pre-populated with data simulating typical warehouse and retail store workloads.

In the live demonstration, we execute the following SASE queries:
- Misplaced inventory query: monitoring for an event where a shelf item appears on the wrong shelf. The detection of such an event triggers an Event Database lookup for the movement history of the item (e.g. all the areas in the store in which the item was previously detected).
- Shoplifting query: monitoring for an event where an item exits the retail store without passing the check-out counter. The stream query performs an Event Database lookup to retrieve a textual description of the exit in which the item left the store.
- Archiving queries: specifying data transformation rules for archiving, *e.g.*, detecting an event representing a change in location of an object and performing a database update to reflect the occurrence this event.

For each of these queries except for archiving queries, we first add the individual query to the complex event processor. Then the actual behavior (e.g. shoplifting or misplaced inventory) is simulated live in our retail store. This results in the real-time detection of the behavior and a notification from the SASE UI. Additionally, the event stream triggers updates of the Event Database according to the database propagation rules registered with the complex event processor. The live updates ensure that all Event Database queries (e.g. ad-hoc or ones triggered by stream queries) are executed over an up-to-date state of the retail store.

For example, consider our demonstration for shoplifting. The query (described in Section 2.1.1) is first registered with the stream processor. Then the shoplifting action is simulated in our retail store, resulting in its successful detection and thus an alert similar to the one in Figure 3. The "Message Results" window (bottom left corner) displays the fully processed output from the shoplifting query specified in the "Present Queries" window (top left corner). The output reflects the result from the database query generated and executed by the *_retrieveLocation(z.AreaId)* function in the RETURN clause joined with *x.TagId*, *x.ProdName*, and *z.AreaId* values computed by the stream query. The three windows in the right-hand column of Figure 3 display the intermediate results computed by SASE. The top right window, "Cleaning and Association Layer Output", monitors the event stream output from the Filtering and Association Layer. The window below, "Database Report", displays the database query triggered by the stream component of the shoplifting query (e.g. the query generated by the _retrieveLocation(z.AreaId) and its result. The "Stream Processor Output" window (below the "Database Report" window) shows the *x.TagId*, *x.ProdName*, and *z.AreaId* values computed by the SASE stream query. The UI uses these intermediate results to provide a useful message to the user.

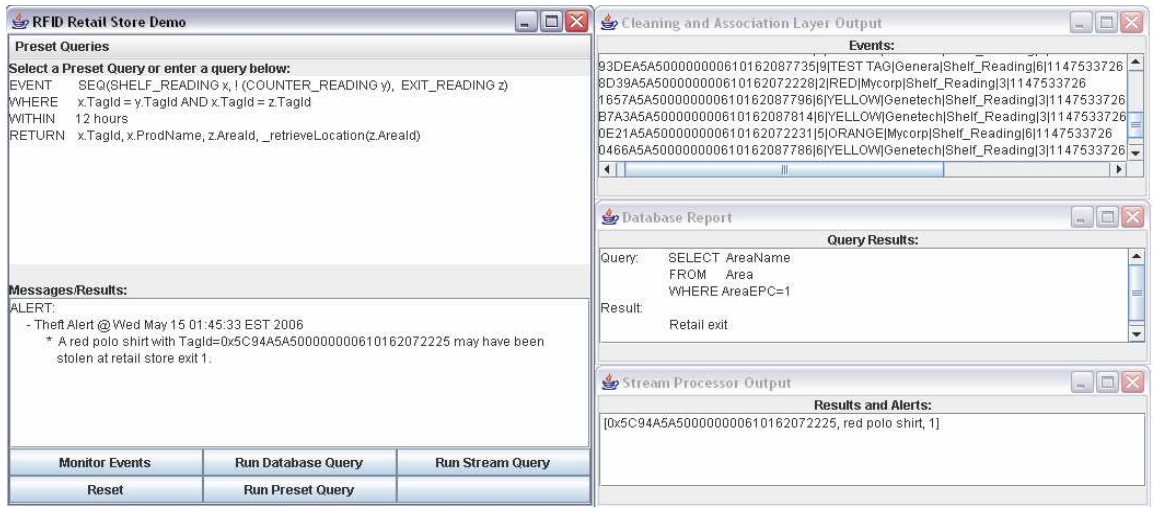

Figure 3: A Screenshot of the SASE UI

During the live demonstration we use the three windows on the right in Figure 3 to demonstrate SASE's internal data flow and display the intermediate results used to compute final query output.

Our demonstration also illustrates the use of track-and-trace queries over the Event Database executed as a part of continuous queries. We pre-populate our Event Database with RFID data that simulates typical warehouse and retail store workloads, such as loading/unloading items, stocking shelves, and changing containments (*e.g.*, moving items from one box to another). This data represents some interesting movement history for our retail-store items throughout a simulated supply chain management system. We run the following track-and trace queries:

- Current location: find the current location of an item.
- Movement history: find the location and containment changes of an item.

Collectively, the live demonstration and track-and-trace queries illustrate the viability of our approach to real-world streaming applications.